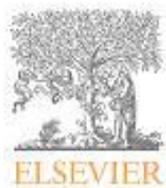
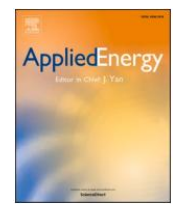

# Decision-focused Conservation Voltage Reduction to Consider the Cascading Impact of Forecast Errors

Qintao Du[a], Ran Li[a,b,*], Weiyi Lv[c], Huan Zhou[a], Moduo Yu[a], Jianzhe Liu[a]

[a]*Key Laboratory of Control of Power Transmission and Conversion, Ministry of Education, and Shanghai Non-Carbon Energy Conversion and Utilization Institute, Shanghai Jiao Tong University, Shanghai 200240, China.*
[b]*College of Smart Energy, Shanghai Jiao Tong University, Shanghai 200240, China.*
[c]*State Grid Shanghai Pudong Electric Power Supply Company, Shanghai 200125, China.*

## HIGHLIGHTS

- Bi-level multi-timescale forecasting framework for multi-stage VVC is proposed.
- Cascading impact of multi-timescale forecast errors on device coordination is mitigated.
- To solve the computationally challenging bi-level model, a sensitivity-driven integer L-shaped solution method is developed.
- Within the proposed solution method, a hybrid gradient feedback mechanism integrating numerical and analytical sensitivities is elaborated.

## ARTICLE INFO

*Keywords:*
Active distribution networks
Conservation voltage reduction
Bi-level optimization
Multi-stage Volt-Var control
Multi-timescale forecasting
Decision-focused learning

## ABSTRACT

Conservation Voltage Reduction (CVR) relies on the effective coordination of slow-acting devices, such as OLTCs and CBs, and fast-acting devices, such as SVGs and PV inverters, typically implemented through a hierarchical multi-stage Volt-Var Control (VVC) spanning day-ahead scheduling, intra-day dispatch, and real-time control. However, existing sequential methods fail to account for the cascading impact of forecast errors on multi-stage decision-making. This oversight results in suboptimal day-ahead schedules for OLTCs and CBs that hinder the effective coordination with fast-acting SVGs and inverters, inevitably driving a trade-off between real-time voltage security and CVR efficiency. To improve the Pareto front of this trade-off, this paper proposes a novel bi-level multi-timescale forecasting (Bi-MTF) framework for multi-stage VVC optimization. By integrating the downstream multi-stage VVC optimization into the upstream forecasting models training, the decision-focused forecasting models are able to learn the trade-offs across temporal horizons. To solve the computationally challenging bi-level formulation, a modified sensitivity-driven integer L-shaped method is developed. It utilizes a hybrid gradient feedback mechanism that integrates numerical sensitivity analysis for discrete variables with analytical dual information for continuous forecast parameters to ensure tractability. Numerical results on a modified IEEE 33-bus system demonstrate that the proposed approach yields superior energy savings and operational safety compared to conventional MSE-based sequential paradigms. Specifically, as the capacity of fast-acting devices increases, the energy savings of the proposed method rise from 2.74% to 3.41%, which is far superior to the 1.50% to 1.76% achieved by conventional MSE-based sequential paradigms.

* Corresponding author.
*E-mail address:* rl272@sjtu.edu.cn (R. Li).

**Nomenclature**

*A. Sets*:

| | |
|---|---|
| $t \in \mathcal{T}$ | Index and set of time periods. |
| $i, j \in \mathcal{N}$ | Index and set of network nodes. |
| $(i, j) \in \mathcal{E}$ | Set of transmission lines connecting node $i$ and $j$. |
| $s \in \mathcal{S}$ | Index and set of historical instances. |
| $m \in \mathcal{M}$ | Index and set of second-order cone constraints. |
| $\mathcal{N}_{PV}, \mathcal{N}_{CB}, \mathcal{N}_{SVG}$ | Subsets of nodes equipped with PV units, Capacitor Banks (CBs), and Static Var Generators (SVGs), respectively. |

*B. Parameters:*

| | |
|---|---|
| $w_L, w_P$ | Weighting coefficients for power losses and load consumptions. |
| $r_{ij}, x_{ij}$ | Resistance and reactance of branch $ij$. |
| $P_{i,t}^{L,rate}, Q_{i,t}^{L,rate}$ | Rated active and reactive load for bus $i$ at time $t$. |
| $Z_p, I_p, P_p$ | Predefined coefficients for active power ZIP load models. |
| $Z_q, I_q, P_q$ | Predefined coefficients for reactive power ZIP load models. |
| $V_{\min}, V_{\max}$ | Minimum and maximum permissible nodal voltage magnitude limits. |
| $V_{base}$ | Base secondary bus voltage at the substation. |
| $\Delta V_{step}$ | Voltage step size of the On-Load Tap Changer (OLTC). |
| $\Delta q_c$ | Reactive power step size of CB $c$. |
| $N_{\max}^{OLTC}, N_{\max,c}^{CB}$ | Predefined maximum daily switching frequencies for the OLTC and CBs. |
| $Q_j^{SVG,\min}, Q_j^{SVG,\max}$ | Minimum and maximum reactive power capacity limits of SVG $j$. |
| $\mathcal{K}$ | Local linear droop coefficients. |
| $\gamma_{DA}, \gamma_{UST}$ | Regularization penalty coefficients for day-ahead and ultra-short-term forecast accuracy. |

*C. Variables:*

| | |
|---|---|
| $\eta^{DA}, \eta^{UST}$ | Trainable forecasting model coefficient vectors for day-ahead and ultra-short-term timescales. |
| $x$ | Vector of discrete schedules (decisions for OLTCs and CBs). |
| $y^{DA}$ | Vector of continuous network state variables in the day-ahead scheduling. |
| $y^{ID}$ | Vector of continuous variables, including the reactive power dispatch of SVGs and updated network states in the intra-day dispatch. |
| $y^{RT}$ | Vector of continuous variables, including the autonomous reactive power responses of PV inverters and the final network states in the real-time control. |
| $P_{i,t}^{PV,DA}, P_{i,t}^{PV,UST}$ | Predicted active power output of PV unit $i$ at time $t$ for the day-ahead and ultra-short-term timescales. |
| $P_{j,t}^{SUB,DA}, Q_{j,t}^{SUB,DA}$ | Active and reactive power injected from the substation at node $j$ at time $t$. |
| $P_{j,t}^{L,DA}, Q_{j,t}^{L,DA}$ | Active and reactive load consumptions at node $j$ at time $t$ based on ZIP models. |
| $v_{i,t}^{DA}$ | Squared nodal voltage magnitude at node $i$ at time $t$. |
| $v_{root,t}^{DA}$ | Squared voltage magnitude at the root node (substation) at time $t$. |
| $l_{ij,t}^{DA}$ | Squared branch current on branch $ij$ at time $t$. |
| $Q_{c,t}^{CB,DA}$ | Reactive power output of CB $c$ at time $t$ in the day-ahead stage. |
| $Q_{i,t}^{SVG,ID}$ | Reactive power output of SVG $i$ at time $t$ in the intra-day dispatch. |
| $g_t^{OLTC}, g_{c,t}^{CB}$ | Integer variables defining the physical tap state of the OLTC and switching step of CB $c$ at time $t$. |
| $\delta_t^{OLTC}, \delta_{c,t}^{CB}$ | Auxiliary variables representing the switching actions of the OLTC and CB $c$ at time $t$. |
| $\theta_s$ | Auxiliary variable representing the lower bound approximation of expected recourse cost in the master problem for scenario $s$. |
| $\lambda_s$ | Optimal dual multipliers associated with the continuous power balance equations for scenario $s$. |
| $Q(x, \omega_s)$ | Recourse function representing the final operational cost under realized uncertainty $s$. |
| $\pi^x, \pi^p$ | Vectors of discrete physical sensitivities and continuous forecast sensitivities. |

## 1. Introduction

Conservation Voltage Reduction (CVR) has emerged as a cost-effective energy-saving technology, which reduces load consumption by lowering the service voltage within permissible limits [1][2]. In active distribution networks, CVR is typically implemented through Volt/Var Control (VVC) [3], utilizing voltage regulation devices that can be categorized as slow-acting devices, such as on-load tap changers (OLTCs) and capacitor banks (CBs), and fast-acting devices, such as static var generators (SVGs) and PV inverters. Specifically, slow-acting devices are controlled via switching on/off and they are usually operated at a limited frequency due to their service lifetime and manufacture techniques.

To effectively coordinate different voltage regulation devices, sequential hierarchical control frameworks have become the standard paradigm. One common implementation adopts a two-stage "optimization-plus-control" approach. In the first stage, a global optimization is performed to schedule slow-acting devices based on day-ahead (DA) forecasts, aiming to minimize total power losses or operational costs. Subsequently, the second stage relies on local control rules, such as Volt-Var or Watt-Var droop controls, to autonomously adjust fast-acting devices in real-time [4]–[7]. Given the significant errors inherent in day-ahead forecasts, recent advancements have adopted multi-timescale forecasting by integrating ultra-short-term (UST) forecasts. This evolves the hierarchy into a multi-stage decision-making comprising day-ahead scheduling, intra-day rolling dispatch, and real-time control [8]-[10]. In this framework, the first two stages typically employ coordinated optimization methods, such as stochastic programming [11]-[13], robust optimization [14][15], chance-constrained optimization [16][17], or model predictive control [18]-[20]. Specifically, based on day-ahead and ultra-short-term forecasts respectively, the first two stages sequentially determine the optimal schedules for slow-acting devices such as OLTCs and CBs, and the dispatch strategies for fast-acting SVGs. Finally, the third stage utilizes local control rules to rapidly fine-tune fast-acting devices, namely PV inverters, in real time to compensate for momentary fluctuations [21][22].

Despite these varied implementations, they share a fundamental limitation in that different forecast error result in different day-ahead schedules for slow-acting devices, which leads to varying real-time operating conditions for fast-acting devices. This inevitably hinders their effective coordination, driving them to trade CVR efficiency for voltage security, or vice versa. Taking a typical sequential framework as an example, an overestimation of day-ahead generation prompts an aggressive schedule for slow-acting devices (i.e., lower OLTC tap). When confronted with the lower-than-expected actual generation, this schedule necessitates substantial real-time reactive power support from fast-acting devices to prevent undervoltage risks, ultimately prioritizing CVR efficiency at the expense of voltage security. Conversely, an underestimation yields a conservative day-ahead schedule for slow-acting devices. Although this minimizes real-time support requirements, it results in a higher average voltage profile, sacrificing CVR efficiency.

To bridge the gap between sequential forecasting and scheduling, the paradigm of decision-focused learning offers a promising pathway by aligning forecasting models directly with downstream operational objectives [23]-[26]. Inspired by this paradigm, to improve the Pareto front of the trade-off between CVR efficiency and voltage security, we propose a novel bi-level multi-timescale forecasting (Bi-MTF) framework for multi-stage VVC optimization. The core innovation lies in considering the cascading impact of forecast errors on device coordination by integrating the downstream multi-stage VVC into the training objective of the upstream multi-timescale forecasting. The framework enables the model to learn the downstream trade-offs, producing tailored multi-timescale forecasts that lead to reliable and cost-effective co-ordination between slow-acting and fast-acting devices.

The key contributions of this paper are as follows:

1) A novel bi-level learning framework is proposed to co-optimize CVR efficiency and voltage security based on multi-timescale forecasting. By embedding the economic and security outcomes of the downstream multi-stage VVC directly into the upstream training objective, the framework creates a decision-focused forecasting model that considers the cascading impacts of multi-timescale forecast errors on device coordination.

2) To solve the computationally challenging bi-level learning formulation, which incorporates the joint training of multi-timescale forecast parameters and mixed-integer downstream recourse decisions, a sensitivity-driven integer L-shaped method is developed. Its key innovation lies in a hybrid gradient feedback mechanism for generating effective optimality cuts. This strategy combines numerical sensitivity analysis for discrete variables via a continuous representation technique with analytical dual information for continuous forecast parameters. This hybrid mechanism efficiently enables a tractable and scalable solution for the complex bi-level learning framework.

## 2. Problem statement

The section introduces the typical sequential Volt/Var control in the active distribution networks, composed of two steps: 1) an upstream multi-timescale PV forecasting process, generating both day-ahead and ultra-short-term forecasts based on contextual features; and 2) a downstream three-stage VVC optimization driven by these forecasts. Specifically, the optimization model determines the optimal day-ahead schedules for slow-acting devices (e.g., OLTCs and CBs) using day-ahead forecasts (Stage 1). Subsequently, it determines the intra-day reactive power dispatch for fast-acting SVGs utilizing updated ultra-short-term forecasts (Stage 2). Finally, the framework accounts for the real-time autonomous reactive power responses of PV inverters (Stage 3) against momentary power fluctuations. The overarching objective across all stages is to minimize the total daily power losses and load consumption.

### 2.1 Upstream multi-timescale PV forecast models

Considering a distribution network with multiple PV systems integrated into various buses, the typical multi-stage voltage regulation strategy relies on PV generation forecasts across different timescales. Specifically, we consider forecasting models for two distinct stages: day-ahead and ultra-short-term. For each bus $i \in \mathcal{N}_{PV}$ equipped with a PV unit at time step $t \in \mathcal{T}$, the forecast generation is formulated as follows:

$$P_{i,t}^{PV,DA} = \Phi_{DA}(\xi_{i,t}; \eta_i^{DA}) \tag{1}$$

$$P_{i,t}^{PV,UST} = \Phi_{UST}(\xi_{i,t}; \eta_i^{UST}) \tag{2}$$

where $P_{i,t}^{PV,DA}$ and $P_{i,t}^{PV,UST}$ represent the predicted active power output of the PV unit at bus $i$ during time interval $t$ for the day-ahead and ultra-short-term timescales, respectively. The function $\Phi(\cdot)$ represents the forecasting model which maps the feature context $\xi_{i,t}$ to the forecast value. The models are parameterized by the trainable vectors $\eta_i^{DA}$ and $\eta_i^{UST}$.

### 2.2 Downstream multi-stage CVR formulation

#### 2.2.1 Stage 1: Day-ahead schedule for slow-acting devices

Based on day-ahead PV forecasts, the day-ahead problem determines the optimal schedule for slow-acting devices, such as OLTCs and CBs. To ensure operational longevity, these devices are governed by constraints on their total number of daily switching operations. The objective function (3) aims to minimize the total load consumptions and power losses using a weighted-sum formulation.

$$\min_{g_t^{OLTC}, g_{c,t}^{CB}} \sum_{t \in \mathcal{T}} \left( w_L \sum_{ij \in \mathcal{E}} r_{ij} l_{ij,t}^{DA} + w_P \sum_{i \in \mathcal{N}} P_{i,t}^{L,DA} \right) \quad (3)$$

$$P_{i,t}^{SUB,DA} + P_{i,t}^{PV,DA} = P_{i,t}^{L,DA} + \sum_{j:ij \in \mathcal{E}} P_{ij,t}^{DA} - \sum_{k:ki \in \mathcal{E}} (P_{ki,t}^{DA} - r_{ki} l_{ki,t}^{DA}),$$
$$\forall i \in \mathcal{N}, \forall t \in \mathcal{T} \quad (4)$$

$$Q_{i,t}^{SUB,DA} + Q_{i,t}^{CB,DA} = Q_{i,t}^{L,DA} + \sum_{j:ij \in \mathcal{E}} Q_{ij,t}^{DA} - \sum_{k:ki \in \mathcal{E}} (Q_{ki,t}^{DA} - x_{ki} l_{ki,t}^{DA}),$$
$$\forall i \in \mathcal{N}, \forall t \in \mathcal{T} \quad (5)$$

$$v_{j,t}^{DA} = v_{i,t}^{DA} - 2(r_{ij} P_{ij,t}^{DA} + x_{ij} Q_{ij,t}^{DA}) + (r_{ij}^2 + x_{ij}^2) l_{ij,t}^{DA},$$
$$\forall ij \in \mathcal{E}, \forall t \in \mathcal{T} \quad (6)$$

$$\left\| \begin{array}{c} 2 P_{ij,t}^{DA} \\ 2 Q_{ij,t}^{DA} \\ l_{ij,t}^{DA} - v_{i,t}^{DA} \end{array} \right\|_2 \leq l_{ij,t}^{DA} + v_{i,t}^{DA}, \quad \forall ij \in \mathcal{E}, \forall t \in \mathcal{T} \quad (7)$$

$$P_{i,t}^{L,DA} = P_{i,t}^{L,rate}(Z_p v_{i,t}^{DA} + I_p \sqrt{v_{i,t}^{DA}} + P_p), \quad \forall i \in \mathcal{N}, \forall t \in \mathcal{T} \quad (8)$$

$$Q_{i,t}^{L,DA} = Q_{i,t}^{L,rate}(Z_q v_{i,t}^{DA} + I_q \sqrt{v_{i,t}^{DA}} + P_q), \quad \forall i \in \mathcal{N}, \forall t \in \mathcal{T} \quad (9)$$

$$V_{\min}^2 \leq v_{i,t}^{DA} \leq V_{\max}^2, \quad \forall i \in \mathcal{N}, \forall t \in \mathcal{T} \quad (10)$$

$$v_{root,t}^{DA} = (V_{base} + g_t^{OLTC} \cdot \Delta V_{step})^2, \quad \forall t \in \mathcal{T} \quad (11)$$

$$-M \delta_t^{OLTC} \leq g_{t+1}^{OLTC} - g_t^{OLTC} \leq M \delta_t^{OLTC}, \quad \forall t \in \mathcal{T} \quad (12)$$

$$\sum_{t \in \mathcal{T}} \delta_t^{OLTC} \leq N_{\max}^{OLTC} \quad (13)$$

$$Q_{c,t}^{CB,DA} = g_{c,t}^{CB} \cdot \Delta q_c, \quad \forall c \in \mathcal{N}_{CB}, \forall t \in \mathcal{T} \quad (14)$$

$$-M \delta_{c,t}^{CB} \leq g_{c,t+1}^{CB} - g_{c,t}^{CB} \leq M \delta_{c,t}^{CB}, \quad \forall c \in \mathcal{N}_{CB}, \forall t \in \mathcal{T} \quad (15)$$

$$\sum_{t \in \mathcal{T}} \delta_{c,t}^{CB} \leq N_{\max,c}^{CB}, \quad \forall c \in \mathcal{N}_{CB} \quad (16)$$

$$\delta_t^{OLTC}, \delta_{c,t}^{CB} \in \{0,1\}, \quad g_t^{OLTC}, g_{c,t}^{CB} \in \mathbb{Z} \quad (17)$$

Constraints (4)–(7) constitute the second-order cone programming (SOCP) relaxation of the DistFlow model [23], incorporating the nodal power balance (4)–(5) and the second-order cone relaxation (6)–(7). (8)-(9) formulates the active and reactive ZIP load models, where $P_{i,t}^{L,rate}$ and $Q_{i,t}^{L,rate}$ are the rated active and reactive load for bus $i$ at time $t$. Constraints (10) enforces the nodal voltage magnitude limits. The operation of slow-acting devices is governed by (11)-(17), which define the physical states of OLTCs and CBs and limit their daily switching frequencies to the predefined maximums $N_{\max}^{OLTC}$ and $N_{\max,c}^{CB}$, respectively.

Mathematically, this day-ahead optimization can be formulated as the following general problem:

$$\min_{x, y^{DA}} f^{DA}(x, y^{DA}) \quad (18)$$

Subject to:

$$A^{DA} x \leq b^{DA} \quad (19)$$

$$E^{DA} y^{DA} = u^{DA} \quad (20)$$

$$\| Q_m^{DA} y^{DA} + q_m^{DA} \|_2 \leq (c_m^{DA})^T y^{DA} + d_m^{DA}, \forall m \in \mathcal{M} \quad (21)$$

$$D^{DA} y^{DA} = g^{DA} - G^{DA} x \quad (22)$$

where $x \in \mathbb{Z}^n$ represents the discrete schedules (OLTCs and CBs). $y^{DA}$ represents the continuous network state variables (nodal voltage, branch currents, and power flows) under the day-ahead forecast. $\mathcal{M}$ denotes the index set of second-order cone constraints.

#### 2.2.2 Stage 2: Intra-day dispatch for fast-acting SVGs

With the day-ahead schedule $x$ for OLTCs and CBs fixed, the second stage performs intra-day (ID) dispatch. This stage introduces the continuous fast-acting SVGs and utilizes ultra-short-term forecasts ($P_{i,t}^{UST}$) to compensate for day-ahead forecast errors and refine voltage profiles. The constraints in Stage 2 are similar to those in Stage 1, including (4), (6)–(10). Key modifications are made to reflect the shift to intra-day dispatch by updating the reactive power balance equations in (25)–(26) to include the reactive power output of SVGs as decision variables. Correspondingly, the OLTC and CB settings are treated as fixed parameters determined in the first stage. Constraints linking between these two stages is given in (27) to ensure that the voltage magnitude at the secondary bus of OLTC remains the same as that at Stage 1, i.e., the OLTC tap switching is scheduled at Stage 1.

$$\min_{Q_{i,t}^{SVG}} \sum_{t \in \mathcal{T}} ( w_L \sum_{ij \in \mathcal{E}} r_{ij} l_{ij,t}^{ID} + w_P \sum_{i \in \mathcal{N}} P_{i,t}^{L,ID} ) \quad (23)$$

Subject to:

Standard DistFlow Constraints (4)-(10) $\quad (24)$

$$Q_{i,t}^{SUB,ID} + Q_{i,t}^{CB,DA} + Q_{i,t}^{SVG,ID} = Q_{i,t}^{L,ID} + \sum_{j:ij \in \mathcal{E}} Q_{ij,t}^{ID}$$
$$- \sum_{k:ki \in \mathcal{E}} (Q_{ki,t}^{ID} - x_{ki} l_{ki,t}^{ID}), \quad \forall i \in \mathcal{N}, \forall t \in \mathcal{T} \quad (25)$$

$$Q_i^{SVG,\min} \leq Q_{i,t}^{SVG,ID} \leq Q_i^{SVG,\max}, \quad \forall i \in \mathcal{N}_{SVG}, \forall t \in \mathcal{T} \quad (26)$$

$$v_{root,t}^{ID} = v_{root,t}^{DA} \quad (27)$$

Mathematically, this intra-day dispatch can be represented in the following general problem:

$$\min_{y^{ID}} f^{ID}(y^{ID}) \quad (28)$$

Subject to:

$$A^{ID} y^{ID} \leq b^{ID} \quad (29)$$

$$\boldsymbol{E}^{ID}\boldsymbol{y}^{ID} = \boldsymbol{u}^{ID} \tag{30}$$

$$\| \boldsymbol{Q}_m^{ID}\boldsymbol{y}^{ID} + \boldsymbol{q}_m^{ID}\|_2 \leq (\boldsymbol{c}_m^{ID})^T \boldsymbol{y}^{ID} + d_m^{ID}, \quad \forall m \in \mathcal{M} \tag{31}$$

$$\boldsymbol{D}^{ID}\boldsymbol{y}^{ID} = \boldsymbol{g}^{ID} - \boldsymbol{G}^{ID}\boldsymbol{x} \tag{32}$$

where $\boldsymbol{y}^{ID}$ represents the continuous variables, including both the reactive power dispatch of SVGs and the updated network state variables in the intra-day dispatch.

*2.2.3 Stage 3: Real-time autonomous control*

In the final stage, while the intra-day reactive power dispatch of SVGs ($Q_{i,t}^{SVG,ID}$) remains fixed, smart PV inverters autonomously provide reactive power support to compensate for momentary active power fluctuations in the real-time (RT) stage. This autonomous adjustment is governed by a local Watt-Var droop control. As demonstrated in [27], the real-time PV reactive power can be parameterized as a linear or piecewise-linear function of active power variations. Following this philosophy, our formulation specializes the local control signal to the ultra-short-term active power forecast error:

$$Q_{i,t}^{PV,RT} = \mathcal{K}_i \cdot (P_{i,t}^{PV,\text{act}} - P_{i,t}^{PV,UST}), \quad \forall i \in \mathcal{N}_{PV}, \forall t \in \mathcal{T} \tag{33}$$

$$Q_{i,t}^{SUB,RT} + Q_{i,t}^{CB,DA} + Q_{i,t}^{SVG,ID} + Q_{j,t}^{PV,RT} = Q_{i,t}^{L,ID} + \sum_{j:ij\in\mathcal{E}} Q_{ij,t}^{RT}$$

$$- \sum_{k:ki\in\mathcal{E}} (Q_{ki,t}^{RT} - x_{ki}l_{ki,t}^{RT}), \quad \forall i \in \mathcal{N}, \forall t \in \mathcal{T} \tag{34}$$

where $Q_{i,t}^{PV,RT}$ denotes the autonomous reactive power response of the PV unit. $P_{i,t}^{PV,\text{act}}$ is the actual realized active power generation, and $\mathcal{K}_i$ represents the predefined local Watt-Var droop coefficient. Crucially, the remaining physical state variables $\boldsymbol{y}^{RT}$ (e.g., nodal voltages and branch flows) are subsequently determined by evaluating the physical operating conditions under the autonomous PV responses $Q_{i,t}^{PV,RT}$. Consistent with the objectives of the preceding stages, the final operational cost $f^{RT}$ quantifies the total load consumption and power losses.

To ensure global optimality across this hierarchical control architecture, we integrate the three stages into a unified mathematical framework. The goal is to determine an optimal first-stage schedule $\boldsymbol{x}$ that minimizes the expected final operational cost over all possible uncertainty scenarios $\omega$, formulated as $\min_{\boldsymbol{x}} \mathbb{E}_\omega[\mathcal{Q}(\boldsymbol{x},\omega)]$. This cost accounts for the sequential cascade where the day-ahead decision constrains the intra-day dispatch, which in turn sets the baseline for the real-time droop response. Here, $\mathcal{Q}(\boldsymbol{x},\omega)$ is the recourse function representing the overall operational cost. For a specific realization of uncertainty $\omega_s$, this recourse function is computed as $\mathcal{Q}(\boldsymbol{x},\omega_s) = \min_{\bar{\boldsymbol{y}}_s} f^{RT}(\boldsymbol{y}_s^{RT},\omega_s)$.

*2.3 Gap between the multi-timescale forecast and multi-stage VVC optimization*

Existing multi-stage VVC framework typically operate under a standard sequential paradigm. A fundamental limitation of this paradigm lies in the isolation of the upstream multi-timescale forecasting models, which are trained solely to maximize statistical accuracy (e.g., minimizing MSE) while ignoring the downstream operational consequences of forecast errors. This disconnect creates a cascading impact where day-ahead and ultra-short-term forecast errors result in suboptimal day-ahead schedules, which then constrain the flexibility of the intra-day dispatch and ultimately compromise the real-time control. Consequently, the day-ahead schedules force a trade-off between real-time voltage security and CVR efficiency. To improve the Pareto front of this trade-off, we propose a novel bi-level learning framework established upon a decision-focused multi-timescale forecasting model. The core innovation lies in embedding the downstream multi-stage VVC optimization directly into the training objective of the upstream multi-timescale forecasting model. This integration enables the forecasting model to learn the trade-offs between CVR efficiency and voltage security, producing tailored forecasts that lead to cost-effective schedules for voltage regulation devices.

### 3. Methodology

To effectively consider the cascading impact of multi-timescale forecast errors across sequential decision stages, this section presents a bi-level learning framework tailored for the multi-stage VVC framework. By employing a sensitivity-driven integer L-shaped decomposition, this methodology ensures coordination between upstream forecasting and downstream scheduling. Following the presentation of the bi-level optimization formulation, the solution methodology is introduced to address the problem's inherent computational challenges. Specifically, a modified integer L-shaped method is developed, employing sensitivity-based cuts to ensure efficient convergence.

*3.1. Formulation of the bi-level learning framework*

To model this error propagation mechanism, the bi-level learning framework is constructed based on a set of historical instances $s \in \mathcal{S}$. This framework establishes a closed loop between upstream multi-timescale forecasting and downstream multi-stage VVC optimization. For each historical instance $s$, the upper level jointly optimizes the parameters for both day-ahead and ultra-short-term forecasting models, while the lower level models the multi-stage VVC framework (comprising day-ahead scheduling, intra-day dispatch, and real-time control) to evaluate the true operational objective resulting from those forecasts. By embedding this operational objective metric into the upper-level objective, the formulation directly optimizes the forecasts to minimize subsequent operational objectives, thereby shaping the forecasting model to become decision-focused. The complete formulation is presented in (35)-(39).

*Upper Level (Forecasting Model Training)*:

$$\min_{\eta^{DA},\eta^{UST}} \frac{1}{|\mathcal{S}|} \left( \sum_{s\in\mathcal{S}} f^{RT}(\boldsymbol{y}_s^{RT},\omega_s) + \sum_{s\in\mathcal{S}} \begin{pmatrix} \gamma_{DA}\|\boldsymbol{P}_s^{DA} - \boldsymbol{P}_s^{act}\|_2^2 + \\ \gamma_{UST}\|\boldsymbol{P}_s^{UST} - \boldsymbol{P}_s^{act}\|_2^2 \end{pmatrix} \right) \tag{35}$$

Subject to:

$$\boldsymbol{P}_s^{DA} = \Phi_{DA}(\xi_s; \eta^{DA}), \quad \boldsymbol{P}_s^{UST} = \Phi_{UST}(\xi_s; \eta^{UST}), \forall s \in \mathcal{S} \quad (36)$$

*Lower Level, Stage 1 (Day-Ahead Scheduling):*

$$\boldsymbol{x}_s \in \arg\min_{\boldsymbol{x}} \ f^{DA}(\boldsymbol{x}; \boldsymbol{P}_s^{DA}), \quad \forall s \in \mathcal{S} \quad (37)$$

*Lower Level, Stage 2 (Intra-Day Dispatch):*

$$\boldsymbol{y}_s^{ID} = \arg\min_{\boldsymbol{y}} f^{ID}(\boldsymbol{y}; \boldsymbol{x}_s, \boldsymbol{P}_s^{UST}), \quad \forall s \in \mathcal{S} \quad (38)$$

*Lower Level, Stage 3 (Real-Time Control):*

$$\boldsymbol{Q}_s^{PV,RT} = \mathcal{K} \cdot (\boldsymbol{P}_s^{PV,act} - \boldsymbol{P}_s^{PV,UST}), \quad \forall s \in \mathcal{S} \quad (39)$$

The upper level (35)-(36) trains the forecasting model parameters $\eta$, i.e., $\eta^{DA}$ and $\eta^{UST}$. The objective function (35) is to find the optimal forecasting model parameters that minimize the total operational objective, which is the sum of the final real-time operational cost $f^{RT}$ and forecast accuracy regularization term to mitigate overfitting. The lower level (37)-(39) serves as a simulation of the multi-stage operational cascade. The first stage (37) models the day-ahead scheduling, which relies on the day-ahead forecast $\boldsymbol{P}_s^{DA}$ to determine optimal discrete schedules for slow-acting devices. The second stage (38) represents the intra-day dispatch, where the continuous reactive power setpoints for fast-acting SVGs are optimized based on ultra-short-term forecasts $\boldsymbol{P}_s^{UST}$. Finally, the third stage (39) represents the real-time autonomous control, which determines the reactive power responses of fast-acting PV inverters via the droop control law $\mathcal{K}$ and evaluates the operational cost $f^{RT}$ under the realized uncertainty $\omega_s$.

### 3.2. Sensitivity-driven integer L-shaped solution methodology

The proposed bi-level learning framework poses significant computational challenges arising from two distinct aspects. First, the complex interdependency between multi-time-scale forecasting and multi-stage VVC framework creates a nested structure. The trainable parameters of both day-ahead and ultra-short-term forecasting model impact the first-stage schedule, which in turn propagates through intra-day dispatch to real-time control. This cascading dependency not only makes the exact evaluation of the recourse function computationally intractable but also complicates the backpropagation of gradients from the final operational objective to the forecasting parameters. Second, the presence of discrete first-stage variables (e.g., OLTC taps and capacitor banks) introduces non-convexity and non-differentiability. This structural characteristic prevents the derivation of standard analytical gradients for the upper-level optimization, resulting in a lack of direct directional guidance for the search process.

To address these challenges, we propose a sensitivity-driven integer L-shaped method. Instead of tackling the intractable bi-level framework directly, this approach decouples the problem into a master problem and operational subproblems. Crucially, it incorporates a hybrid gradient feedback mechanism into the iterative process, which effectively backpropagates downstream operational sensitivities to guide the joint optimization of discrete schedules and forecast parameters.

#### 3.2.1 Master problem formulation

The master problem (MP) serves as the centralized decision-maker, jointly optimizing the discrete day-ahead schedule $x$ and the forecasting model parameters $\eta$. To enable efficient gradient-based search within the operational space, we propose a hybrid gradient feedback mechanism. This strategy combines numerical sensitivity analysis for discrete variables, achieved via a continuous representation technique [24], with analytical dual information for continuous forecast parameters. This hybrid mechanism enables the generation of valid sensitivity-based cuts that jointly map the operational cost gradients to both the physical decision space and the forecast parameter space. The complete master problem is presented in (40)-(48):

$$MP : \min_{\boldsymbol{x}_s, \eta, \theta_s} \sum_{s \in \mathcal{S}} p_s \theta_s + \sum_{s \in \mathcal{S}} \begin{pmatrix} \gamma_{DA} \| \boldsymbol{P}_s^{DA} - \boldsymbol{P}_s^{act} \|_2^2 + \\ \gamma_{UST} \| \boldsymbol{P}_s^{UST} - \boldsymbol{P}_s^{act} \|_2^2 \end{pmatrix} \quad (40)$$

Subject to:

$$\boldsymbol{P}_s^{DA} = (\eta^{DA})^T \xi_s, \quad \boldsymbol{P}_s^{UST} = (\eta^{UST})^T \xi_s, \forall s \in \mathcal{S} \quad (41)$$

$$\boldsymbol{A}^{DA} \boldsymbol{x}_s \leq \boldsymbol{b}^{DA}(\boldsymbol{P}_s^{DA}), \forall s \in \mathcal{S} \quad (42)$$

$$\boldsymbol{E}^{DA} \boldsymbol{y}_s^{DA} = \boldsymbol{u}^{DA}(\boldsymbol{P}_s^{DA}), \forall s \in \mathcal{S} \quad (43)$$

$$\| \boldsymbol{Q}_m^{DA} \boldsymbol{y}_s^{DA} + \boldsymbol{q}_m^{DA} \|_2 \leq (\boldsymbol{c}_m^{DA})^T \boldsymbol{y}_s^{DA} + d_m^{DA}, \forall s \in \mathcal{S}, \forall m \in \mathcal{M}$$

$$(44)$$

$$\boldsymbol{x}_s \in \mathbb{Z}^n, \forall s \in \mathcal{S} \quad (45)$$

$$r_s^{OLTC} = V_{base} + g_s^{OLTC} \cdot \Delta V_{step}, \forall s \in \mathcal{S} \quad (46)$$

$$r_{s,c}^{CB} = g_{s,c}^{CB} \cdot \Delta q_c, \forall s \in \mathcal{S}, \forall c \in \mathcal{N}_{CB} \quad (47)$$

$$\theta_s \geq \mathcal{Q}_s^{(k)} + \underbrace{(\pi_s^{x,(k)})^T (\boldsymbol{r}_s - \boldsymbol{r}_s^{(k)})}_{\text{Discrete Sensitivity}}$$
$$+ \underbrace{(\pi_s^{P,(k)})^T (\boldsymbol{P}_s^{UST} - \boldsymbol{P}_s^{UST,(k)})}_{\text{Forecast Sensitivity}}, \forall s \in \mathcal{S} \quad (48)$$

where (40) minimizes the total expected cost and forecast accuracy regularization terms. (41) adopts a linear forecasting formulation to maintain computational tractability, as embedding complex non-linear structures, such as deep neural networks, into the optimization would render the problem computationally prohibitive [25]. Based on the day-ahead forecast $\boldsymbol{P}_s^{DA}$, (42)–(45) enforce the first-stage constraints on the integer variables $\boldsymbol{x}_s$. (46) and (47) define the continuous representation, where $\boldsymbol{r}$ serves as a unified continuous counterpart representing the OLTC tap ratios and the reactive power injections of CBs, with $\Delta V_{step}$ and $\Delta q_c$ denoting their respective physical step sizes. (48) integrates the discrete sensitivity $\pi^x$ and the forecast sensitivity $\pi^p$ into a valid supporting hyperplane, providing the effective hybrid gradient feedback to guide the optimization and forecasting model training.

To computationally derive these sensitivity coefficients $\pi^x$ and $\pi^p$, we combine finite difference [28] for the discrete physical variables with analytical duality for the continuous forecast parameters [29].

*a) Finite Difference for Discrete Variables* ($\pi^x$): Due to the non-differentiable nature of discrete scheduling, analytical gradients cannot be directly derived. Therefore, we utilize a forward finite-difference method to compute each element of the discrete sensitivity $\pi^x$. By perturbing the decision along the standard unit vector $e_j$ (physically corresponding to a single-step shift, e.g., $\Delta V_{step}$ for OLTC) and re-evaluating the subproblem, we numerically approximate the directional gradient with respect to the equivalent continuous physical variable $r$:

$$\pi_{s,j}^{x,(k)} \approx \frac{\mathcal{Q}(x_s^{(k)} + e_j, P_s^{UST,(k)}) - \mathcal{Q}(x_s^{(k)}, P_s^{UST,(k)})}{\Delta_j} \quad (49)$$

  *b) Dual Variables for Continuous Forecasts* ($\pi^p$): In contrast to the discrete variables, the ultra-short-term forecast $P^{UST}$ enters the convex SOCP subproblem as a continuous parameter within the linear power balance constraints. Consequently, we leverage analytical duality to retrieve the exact gradient from the optimal dual multipliers $\lambda_s$ associated with the power balance equations. According to the sensitivity analysis for convex optimization, this gradient is given by:

$$\pi_s^{p,(k)} = \frac{\partial \mathcal{Q}}{\partial P_s^{UST}} = \lambda_s \quad (50)$$

**Proposition 1**. *The set of inequalities in (48), utilizing the proposed hybrid sensitivity coefficients, yields a set of valid optimality cuts.*

**Proof**. The validity is established by demonstrating that the cut provides a valid lower bound for the final operational objective function $\mathcal{Q}$. The cut functions as a supporting hyperplane of the cost function constructed at the historical iteration point $(r_s^{(k)}, P_s^{UST,(k)})$. We consider two primary cases:

1. If the current solution coincides with the historical iteration point (i.e., $r_s = r_s^{(k)}$ and $P_s^{UST} = P_s^{UST,(k)}$), both gradient summation terms in the cut (12.9) vanish identically. The inequality simplifies to $\theta_s \geq \mathcal{Q}_s^{(k)}$, which creates a tight bound at the current solution.
2. If the current solution deviates from the historical iteration point (i.e., $r_s \neq r_s^{(k)}$ or $P_s^{UST} \neq P_s^{UST,(k)}$), the right-hand side of the cut provides a linear approximation based on sensitivities. For the discrete component, the finite difference $\pi^x$ provides a valid directional secant approximation of the cost change with respect to physical state changes. For the continuous component, the dual multiplier $\pi^p$ provides the exact sub-gradient of the convex subproblem's value function with respect to forecast parameters.

Collectively, this hybrid approximation constitutes a valid supporting hyperplane that globally bounds $\mathcal{Q}$ from below. Unlike standard integer cuts, this sensitivity-based formulation provides effective directional guidance for unexplored solutions, thereby ensuring superior generalization and significantly accelerating convergence.

### 3.2.2 Sequential subproblem formulation

After the master problem determines a new day-ahead schedule $x_s^{(k)}$ and UST forecast $P_s^{UST,(k)}$, the subproblem (SP) evaluates the operational objective $Q(x_s, \omega_s)$ by simulating the sequential response of the fast-acting devices, i.e., SVGs and smart PV inverters, given the realized uncertainty $\omega_s$. The SP problem is formulated as follows:

*Phase I Intra-Day Dispatch:* Given the fixed day-ahead schedule $x_s$ and the ultra-short-term forecast $P_s^{UST}$, the intra-day dispatch is performed to determine the optimal reactive power setpoints for SVGs. To facilitate the extraction of dual information ($\lambda_s$), we formulate this intra-day dispatch problem as a convex SOCP:

$$\boldsymbol{SP}: \min_{y_s^{ID}} f^{ID}(y_s^{ID}; x_s, P_s^{UST}) \quad (51)$$

Subject to:

$$\boldsymbol{A}^{ID} y_s^{ID} \leq \boldsymbol{b}^{ID} \quad (52)$$
$$\boldsymbol{E}^{ID} y_s^{ID} = \boldsymbol{u}^{ID} \quad (53)$$
$$\boldsymbol{D}^{ID} y_s^{ID} = \boldsymbol{g}^{ID}(P_s^{UST}) - \boldsymbol{G}^{ID} x_s \quad : \lambda_s \quad (54)$$
$$\|\boldsymbol{Q}_m^{ID} y_s^{ID} + \boldsymbol{q}_m^{ID}\|_2 \leq (\boldsymbol{c}_m^{ID})^T y_s^{ID} + d_m^{ID}, \quad \forall m \in \mathcal{M} \quad (55)$$

*Phase II Real-Time Droop Control:* Under the realized scenario $\omega_s$, the autonomous reactive power responses of PV inverters are directly determined by the predefined linear droop mapping $\mathcal{K}$ formulated in (39). The final operational objective is calculated by aggregating the real-time operational cost $f^{RT}$ under $y_s^{RT}$ with soft penalties for any voltage violations to ensure robust feasibility.

$$\mathcal{Q}(x_s, \omega_s) = f^{RT}(y_s^{RT}; x_s, P_s^{act}) + M\boldsymbol{1}^T(\Delta \boldsymbol{V}_s^{over} + \Delta \boldsymbol{V}_s^{under}) \quad (56)$$

Subject to:

$$V_{min}^2 \boldsymbol{1} - \Delta \boldsymbol{V}_s^{under} \leq v_s^{RT} \leq V_{max}^2 \boldsymbol{1} + \Delta \boldsymbol{V}_s^{over} \quad (57)$$
$$\Delta \boldsymbol{V}_s^{over} \geq \boldsymbol{0}, \quad \Delta \boldsymbol{V}_s^{under} \geq \boldsymbol{0} \quad (58)$$

where $\boldsymbol{1}$ and $\boldsymbol{0}$ are column vector of ones and zeros, respectively, and $\Delta \boldsymbol{V}_s^{over}$ and $\Delta \boldsymbol{V}_s^{under}$ are the continuous vectors of over-voltage and under-voltage violations.

The complete offline training procedure is summarized in Algorithm 1. Once the optimal forecasting model parameters $\eta^*$ is obtained, the online operation functions as a computationally efficient sequential stages, directly executing the day-ahead scheduling, intra-day dispatch, and real-time control without iterative searching.

---

**Algorithm 1:** Modified sensitivity-driven integer L-shaped method

**Input:** Training scenarios $\mathcal{S}$, iteration limit $K_{max}$, gap tolerance $\epsilon$.

**Definition:** Let $\mathcal{R}(\eta) = \sum_{s \in \mathcal{S}}(\gamma_{DA}\|P_s^{DA} - P_s^{act}\|_2^2 + \gamma_{UST}\|P_s^{UST} - P_s^{act}\|_2^2)$ denote the forecast penalty.

**Output:** Optimally forecasting model parameters $\eta^*$.

**Initialize:** Index of iterations $k = 0$, lower bound $LB \leftarrow -\infty$, upper bound $UB \leftarrow +\infty$, initial model parameters $\eta^{(0)}$.

**while** $k < K_{max}$ **and** $(UB - LB)/|LB| > \epsilon$ **do**

  Solve the **MP** (40)-(48) obtain $(\eta^{(k+1)}, \{x_s^{(k+1)}\}, \{\theta_s^{(k+1)}\})$.

  Update the lower bound: $LB \leftarrow \sum_{s \in \mathcal{S}} p_s \theta_s^{(k+1)} + \mathcal{R}(\eta^{(k+1)})$.

  **For** each instance $s \in \mathcal{S}$ **do**

   Update forecast $P_s^{UST,(k+1)}$ and solve **SP** (51)-(55) to obtain cost $\mathcal{Q}_s^{(k+1)}$ and duals $\lambda_s$.

   **(a) Discrete Sensitivity:** Compute $\pi_s^{x,(k+1)}$ via finite difference perturbations (49).

   **(b) Forecast Sensitivity:** Set $\pi_s^{p,(k+1)} \leftarrow \lambda_s$ (50).

  **end**

  Calculate the total true cost: $Z^{(k+1)} \leftarrow \sum_{s \in \mathcal{S}} p_s \mathcal{Q}_s^{(k+1)} + \mathcal{R}(\eta^{(k+1)})$.

```
    Update the upper bound:  UB ← min{UB, Z^(k+1)} .
    Calculate the optimality gap:  gap ← (UB − LB)/ | LB | .
    if  gap ≤ ϵ  then
        η* ← η^(k+1), x* ← x^(k+1) ; Break
    else
        Add sensitivity-based cuts (48) to MP using  {Q_s, π_s^x, π_s^p} .
        Update Trust Region center:  η^(k) ← η^(k+1)
        k ← k + 1
    end
end
```

## 4. Case study

### 4.1. Case description

i) Testing system and dataset.

The proposed method is validated on a modified IEEE 33-bus distribution network [30], with the locations of the OLTC, CBs, and SVGs shown in Fig. 1. Multiple PV station data combined with related weather information from 2018 to 2019 are utilized from the NREL National Solar Radiation Database (NSRDB) [31]. The weather data with a 60-minute resolution include temporal indicators (year, month, day, hour, minute) and weather variables including irradiance, air pressure, wind direction, wind speed, temperature, and humidity. As feature selection is not the focus of this work, these commonly used features were selected to form the feature set. To conduct the empirical validation, 80% of the historical data was assigned to train the forecasting models, leaving a 20% holdout set to assess the final operational outcomes. All simulations were implemented in Python and solved using Gurobi on a PC equipped with an Intel i7-13700U CPU and 32 GB RAM.

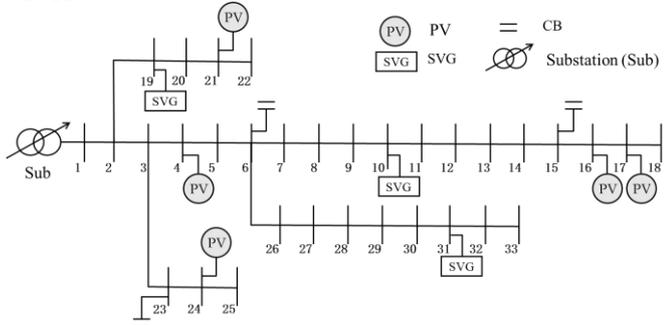

**Fig. 1.** Topology of modified IEEE 33-bus distribution network.

ii) Device and Network Parameters.

An OLTC is located at the substation, regulating the secondary bus voltage within a [0.9, 1.1] p.u. range. It has 21 discrete tap positions (±10 taps plus neutral) with a step size of 0.01 p.u. The network includes three 300 kVAR CBs, each with three 100 kVAR steps, and three SVGs, each with a reactive power range of [-100, 300] kVAR. The voltage-dependent electric loads, connected to all buses except the substation, are modeled using a ZIP formulation with coefficients $[P_Z, P_I, P_P, Q_Z, Q_I, Q_P] = [0.96, -1.17, 1.21, 6.28, -10.16, 4.88]$ [32]. The nominal voltage $V_0$ is 1.0 p.u., and the permissible operational voltage range is maintained within the ANSI standard limits of [0.95, 1.05] p.u.

### 4.2. Convergence results and forecasting performance

#### 4.2.1. Training progress

Fig. 2 evaluates the convergence efficiency of the proposed sensitivity-driven integer L-shaped method compared with the classical L-shaped approach. The classical method (Fig. 2(a)) exhibits poor convergence, as the optimality gap stops to narrow at an early stage, leaving a significant optimality gap. In contrast, the lower bound of the proposed method (Fig. 2(b)), accelerated by the powerful sensitivity-based cuts, rises rapidly to close the optimality gap with the upper bound, reaching convergence at 51.18 within 6 iterations. This confirms the effectiveness of the proposed hybrid gradient feedback mechanism. By integrating numerical sensitivity for discrete variables with analytical duality from continuous forecast parameters, the generated cuts provide precise directional guidance that facilitates efficient convergence.

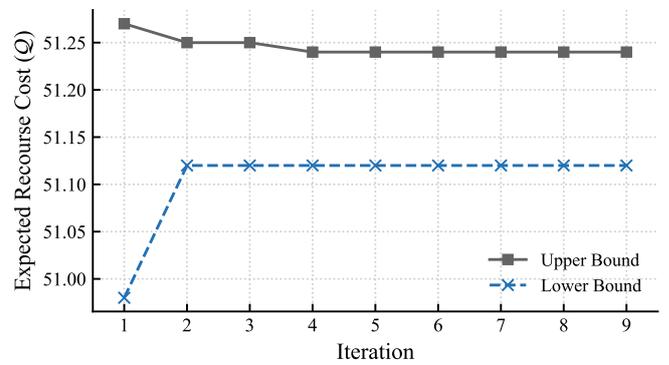

(a) Classical L-shaped method

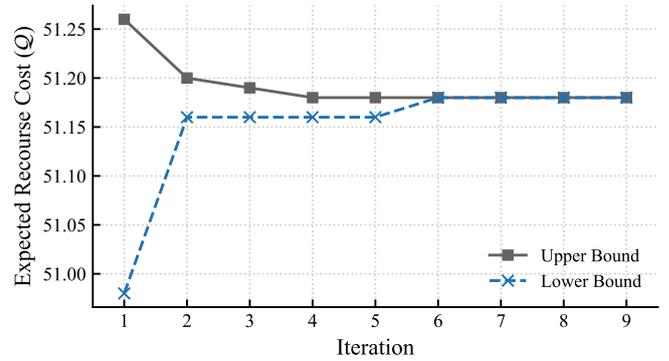

(b) Modified sensitivity-driven integer L-shaped method

**Fig. 2.** Convergence performance comparison between the classical L-shaped method and the proposed sensitivity-driven integer L-shaped method.

#### 4.2.2. Forecasting performance

To evaluate the adaptability of the decision-focused forecasting model to different system flexibility, Table 1 details the normalized RMSE (NRMSE) of the forecasts generated by the conventional MSE-based forecasting method and the proposed Bi-MTF method under varying SVG capacity levels. Correspondingly, Fig. 3 presents the adaptive forecasts generated by the proposed Bi-MTF method under varying levels of system flexibility.

As shown in Table 1, the conventional MSE-based method consistently achieves the lowest NRMSE, as it is optimized solely for statistical accuracy. In contrast, the NRMSE of the

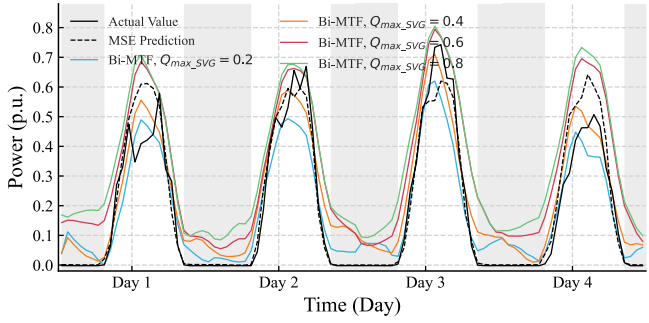

**Fig. 3.** Comparison among the actual PV generation value, the MSE prediction, and the adaptive forecasts generated by the proposed Bi-MTF method under varying SVG capacities.

**Table 1**

Comparison of forecasting performance (NRMSE %) between MSE and BI-MTF methods under varying SVG capacities.

| Level | MSE-based method | Proposed BI-MTF method | | | |
|---|---|---|---|---|---|
| | | 0.2 | 0.4 | 0.6 | 0.8 |
| PV1 | 7.41 | 14.45 | 10.08 | 14.98 | 15.86 |
| PV2 | 7.16 | 13.78 | 9.72 | 14.25 | 14.75 |
| PV3 | 6.54 | 12.45 | 9.15 | 12.94 | 13.42 |
| PV4 | 8.42 | 16.78 | 11.24 | 17.15 | 17.86 |
| PV5 | 7.45 | 14.87 | 10.12 | 15.54 | 15.97 |
| Average | 7.40 | 14.47 | 10.06 | 14.97 | 15.57 |

proposed Bi-MTF method is slightly higher and, notably, exhibits variations correlated with the available SVG capacity. Fig. 3 further illustrates the specific adaptive behaviors. In the shaded nighttime regions, the proposed Bi-MTF method consistently outputs non-zero generation forecasts, which systematically increase in line with the available SVG capacity. During the unshaded daytime regions, the daytime forecasts are adaptively scaled according to the available SVG capacity, ranging from lower forecasts under limited capacity (0.2-0.4 MVar) to higher forecasts under abundant capacity (0.6-0.8 MVar). These adaptive forecasts indicate that the BI-MTF method learns to adjust the forecast based on the system flexibility limits.

### 4.3. Impact of adaptive forecasts on operational performance

We evaluate the operational performance of the proposed BI-MTF method against two baselines: BASE [8] and Oracle (an idealized BASE with perfect information). For these three evaluated methods, Fig. 4 and Fig. 5 illustrate the total energy savings and the total number of voltage violations on the test set, respectively.

The BASE method exhibits limited adaptability and robustness. As shown in Fig. 4, it achieves minimal energy savings from 1.50% to 1.76% with no improvement as SVG capacity increases, indicating an inability to utilize the added flexibility of downstream SVG capacity. Furthermore, as revealed in Fig. 5, the BASE method exposes its fragility under the limited 0.2-MVar SVG capacity, resulting in a substantial 145 voltage violations. In contrast, the proposed BI-MTF method demonstrates strong adaptability. In terms of energy efficiency (Fig. 4), its savings rise significantly from 2.74% to 3.41% as SVG capacity increases. Regarding safety (Fig. 5), it maintains minimal violations across all capacity levels, effectively matching the Oracle's record. Crucially, BI-MTF outperforms even the Oracle in energy savings.

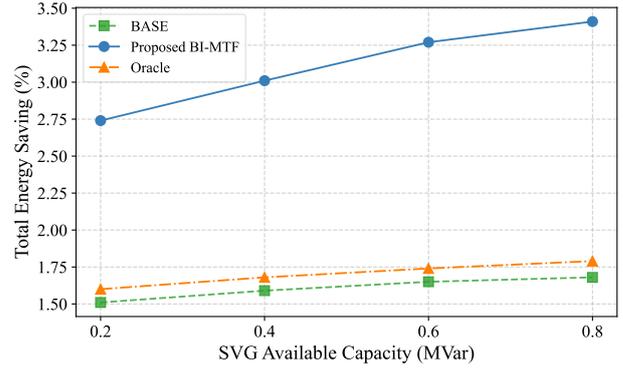

**Fig. 4.** Comparison of total energy saving achieved by the BASE, Proposed BI-MTF, and Oracle methods across varying levels of SVG capacity.

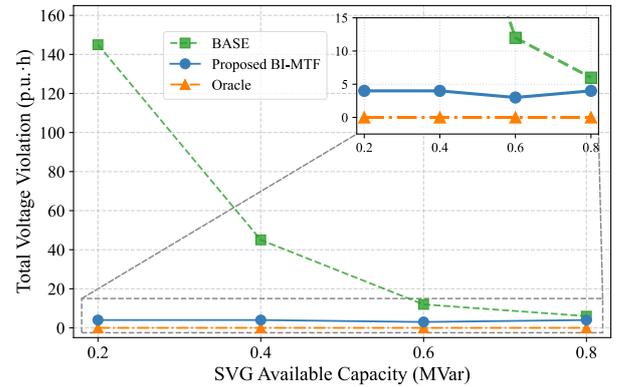

**Fig. 5.** Comparison of total voltage violations resulting from the BASE, Proposed BI-MTF, and Oracle methods across varying levels of SVG capacity.

To investigate the mechanisms behind these operational results via device coordination, we conduct a comparative analysis of the day-ahead schedules and the resulting real-time performance under low (0.2 MVar) and high (0.6 MVar) capacities. Based on the operation day shown in Fig. 6 and Fig.7, we draw the following conclusions on the coordination mechanisms:

*a) The BASE method reveals the limitations of the sequential optimization paradigm:* As shown in Fig. 6(a), its forecasts remain invariant for both 0.2 MVar and 0.6 MVar capacities. This occurs because the MSE method minimizes statistical error independent of the downstream system flexibility provided by SVGs. As observed in Fig. 6(b), these invariant forecasts serve as the fixed input for the day-ahead scheduling, leading to a static and high OLTC schedule across all capacity scenarios. This leads to the behavior observed in Fig. 6(c) and Fig. 6(d), where the SVG operation is restricted to unidirectional reactive power absorption (indicated by negative values), particularly in the shaded nighttime regions. Furthermore, due to the lack of coordination, the method fails to ensure system security under limited SVG capacity (0.2 MVar) during the daytime regions. This results in signific ant voltage violations when large forecast errors occur, notably at 14:00 and 16:00.

*b) The proposed method strategically co-optimizes CVR efficiency and system security by integrating downstream*

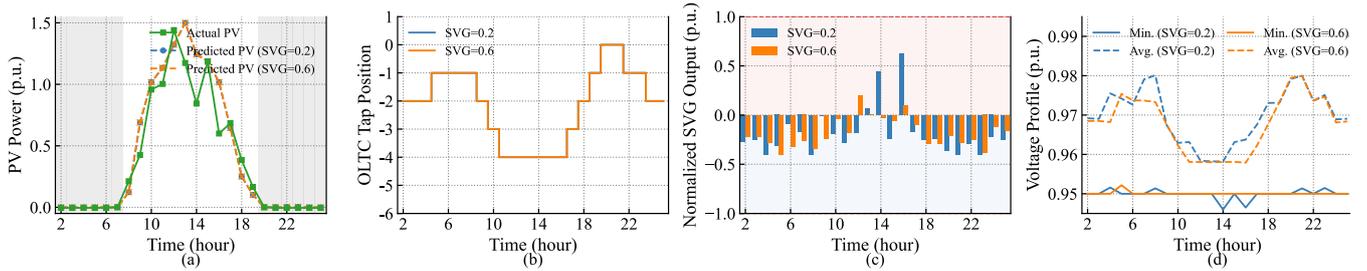

**Fig. 6.** Operational schedules and system performance of the BASE method under varying SVG capacities.

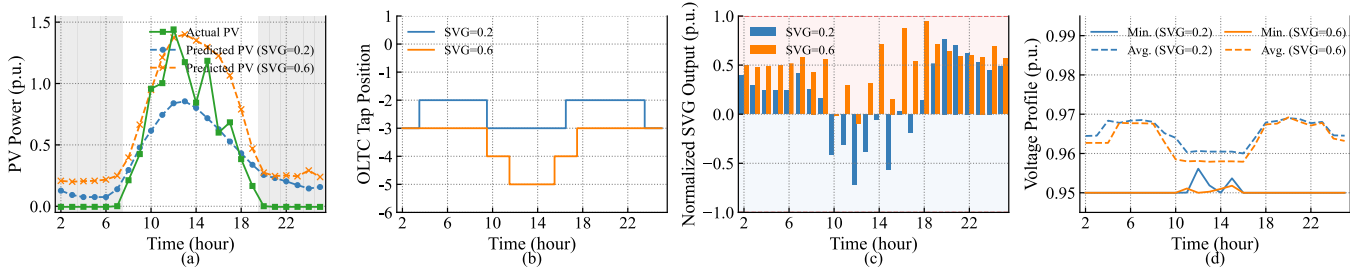

**Fig. 7.** Operational schedules and system performance of the proposed BI-MTF method under varying SVG capacities.

*flexibility into the day-ahead scheduling:* This is achieved through the adaptive day-ahead forecasts, as illustrated in Fig. 7(a), which exhibit adaptability to the available SVG capacity. The method's superior CVR efficiency and security stems from a consistent nighttime strategy and a dynamic daytime strategy.

During the shaded nighttime regions, as shown in Fig. 7(a), the proposed method issues a non-zero generation forecast. As depicted in Fig. 7(b) and Fig. 7(c), this higher forecast drives the model to select a lower OLTC tap position to reduce the global voltage profile, while the SVGs actively inject reactive power (indicated by positive values) to support the voltage at critical nodes to prevent under-voltage violations. This effectively achieves the global coordination that the BASE method missed, thereby maximizing nighttime CVR potential.

During the unshaded daytime regions, the strategy adapts to manage the trade-off between CVR efficiency and voltage security. With a limited 0.2 MVar SVG capacity, the forecast is lower to create a security margin, ensuring a safe OLTC schedule and mitigating security risks. With abundant 0.6 MVar SVG capacity, the daytime forecast becomes higher. As indicated in Fig. 7(b) and Fig. 7(c), this facilitates a lower OLTC schedule and the reactive support of the SVGs is fully exploited, which leads to a lower average voltage profile as evidenced in Fig. 7(d).

*4.4. Analysis of adaptive forecasting distribution and nodal voltage impact*

This section presents an analysis of the adaptive forecasting distribution of the proposed method and their physical impact on the network.

*1) Analysis of Adaptive Forecast Distribution:* Fig. 8 presents scatter plots comparing the forecasts from the proposed method (Y-axis) against those from the MSE method (X-axis) under four SVG capacity levels (0.2, 0.4, 0.6, and 0.8 MVar).

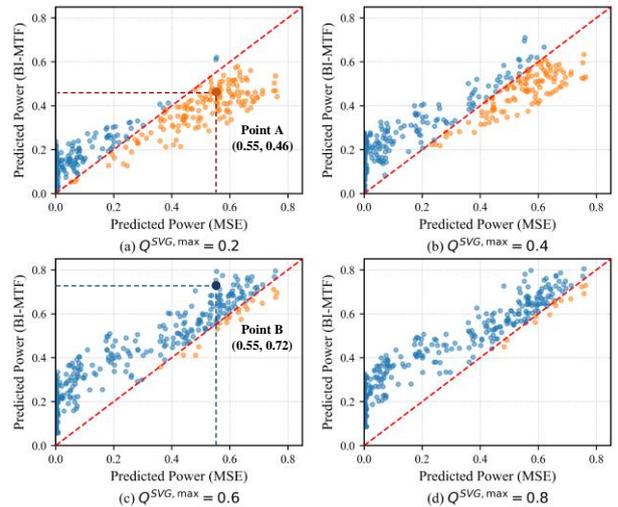

**Fig. 8.** Scatter plots comparing forecasts from the proposed Bi-MTF method (y-axis) and BASE method (x-axis) under four SVG capacity levels (0.2, 0.4, 0.6, and 0.8 MVar). Blue points indicate situations where the Bi-MTF forecast value exceeds that of MSE, while orange points represent the opposite.

The identity red line indicates that x-axis and y-axis approaches have the same forecast value. Blue points indicate situations where the proposed method forecasts a value higher than the MSE method, while vice versa for the orange points.

In the low generation range (e.g., when the MSE forecast is below 0.2 MVar), a consistent strategy is observed across all four subplots. The forecasts from the proposed method are almost universally located above the y=x diagonal line. This indicates that the proposed method consistently generates forecasts higher than the MSE values across all levels of SVG capacity. In the high generation range (e.g., when the MSE forecast exceeds 0.2 MVar), the distribution varies significantly with varying levels of SVG capacity. Under limited SVG capacity, as shown in Fig. 8(a), the data points are pre-dominantly orange, clustering below the identity line. In contrast, as SVG capacity increases, the distribution gradually shifts upward. The data points migrate above the identity line (becoming blue), showing that the forecast values generated by the proposed BI-MTF method increases.

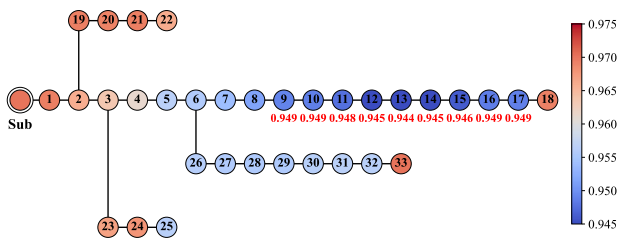

(a) MSE method

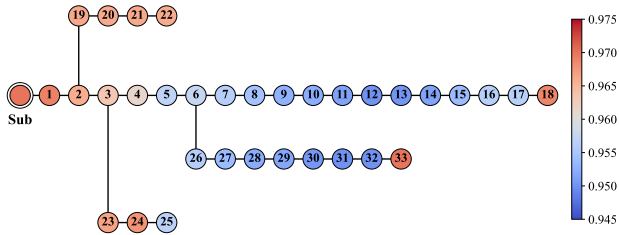

(b) The proposed BI-MTF method

**Fig. 9.** Nodal voltage topology corresponding to Point A (limited SVG capacity, $Q^{SVG,max} = 0.2$MVar). (a) The MSE method leads to undervoltage violations (deep blue nodes). (b) The proposed BI-MTF method successfully mitigates violations and ensures voltage security.

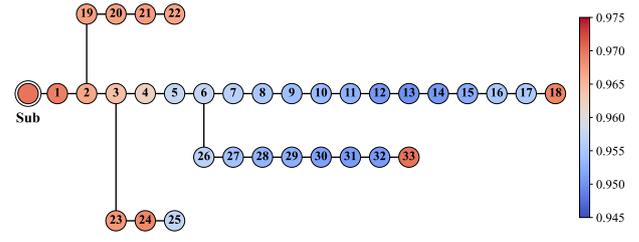

(a) MSE method

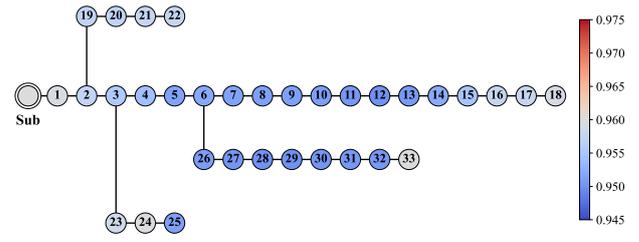

(b) The proposed BI-MTF method

**Fig. 10.** Nodal voltage topology corresponding to Point B (abundant SVG capacity, $Q^{SVG,max} = 0.6$MVar). (a) The BASE method results in a conservative high-voltage profile (red/orange nodes). (b) The proposed BI-MTF method effectively lowers the global voltage profile.

To quantify this adaptive behavior, we analyze two representative points, A and B, where the MSE method yields the same forecast value of 0.55 p.u. At Point A (Fig. 8(a)), under the limited SVG capacity of 0.2 MVar, the proposed method yields a forecast of 0.46 p.u. (Coordinate: 0.55, 0.46). Compared to the MSE method, the proposed method lowers the forecast value. This lower forecast prioritizes a safer OLTC schedule to mitigate undervoltage risks. Conversely, at Point B (Fig. 8(c)) with an abundant capacity of 0.6 MVar, the forecast for the same scenario rises to 0.72 p.u. (Coordinate: 0.55, 0.72), exceeding the value generated by the MSE method. This higher forecast, driven by the abundant downstream flexibility, leads the day-ahead strategy to select a lower OLTC tap to maximize CVR savings.

*2) Nodal Voltage Topology Visualization:* To visualize the direct physical consequences of these adaptive behavior, Fig. 9 and Fig. 10 present the nodal voltage topologies for these two representative points, A and B. Nodes are rendered using a colormap where red indicates high voltage and blue indicates low voltage.

In the limited SVG capacity scenario at Point A, Fig. 9(a) shows that the MSE method's aggressive forecast (0.55 p.u.) leads to a higher OLTC schedule. Due to the insufficient 0.2 MVar SVG capacity, the system fails to manage the resulting voltage profile, causing undervoltage violations at nodes 9-17 (deep blue nodes). In contrast, Fig. 9(b) shows that the proposed method's conservative forecast value (0.46 p.u.) for Point A results in a safer OLTC schedule. This effectively mitigates the undervoltage violations at nodes 9-17, bringing them back to a secure level.

In the abundant SVG capacity scenario at Point B, Fig. 10(b) illustrates that the proposed method's aggressive forecast (0.72 p.u.) successfully enables a lower OLTC tap, as visually evidenced by the substation (Sub) node exhibiting a lower voltage level (grey nodes). This day-ahead OLTC schedule is effectively accommodated by the abundant SVG capacity in real-time, resulting in a globally lower voltage profile (light/dark blue nodes) and achieving maximum CVR efficiency. Conversely, as shown in Fig. 10(a), the MSE method results in an overall high voltage profile (orange/red nodes), particularly at nodes 1-4 and 19-24, failing to exploit the CVR potential offered by the abundant SVG capacity.

## 5. Conclusion

This paper has established a bi-level multi-timescale forecasting framework to co-optimize CVR efficiency and voltage security, effectively resolving the conflicts caused by the separation of multi-timescale forecasting and multi-stage optimization in traditional sequential dispatch. By embedding the outcomes of the downstream multi-stage VVC optimization directly into the upstream forecasting model training, the proposed method internalizes the cascading impact of multi-timescale forecast errors, prioritizing final decision quality over statistical ac-curacy. Crucially, the forecasting model trained by the proposed Bi-MTF framework demonstrates an adaptive strategy driven by system flexibility. Specifically, during nighttime, the model consistently generates aggressive forecasts to drive lower OLTC tap positions. This action reduces the voltage profile while actively leveraging the reactive power support from fast-acting devices to maximize CVR potential. Conversely, during daytime, the model dynamically shifts its forecasts based on system flexibility: it adheres to conservative fore-casts under limited SVG capacity to ensure security, while transitioning to aggressive forecasts under abundant SVG capacity to fully exploit energy savings. These findings con-firm that the proposed method effectively resolves the conflict between efficiency and security, resulting in an overall reduced voltage profile and superior energy savings compared to conventional sequential paradigms.

**CRediT authorship contribution statement**

**Qintao Du:** Writing – original draft, Methodology, Formal analysis, Investigation. **Ran Li:** Methodology, Writing – review & editing, Supervision, Conceptualization. **Weiyi Lv:** Methodology, Conceptualization. **Huan Zhou:** Methodology, Investigation. **Moduo Yu:** Investigation, Writing – review & editing. **Jianzhe Liu:** Methodology, Writing – review & editing.

**Declaration of Competing Interest**

The authors declare that they have no known competing financial interests or personal relationships that could have appeared to influence the work reported in this paper.

**Data availability**

Data will be made available on request.